\documentclass[11pt, a4paper]{article}

\usepackage{amsmath,amssymb}
\usepackage[T1]{fontenc}
\usepackage[english]{babel}

\newcommand{\bea}{\begin{eqnarray}}
\newcommand{\eea}{\end{eqnarray}}
\newcommand{\beano}{\begin{eqnarray*}}
\newcommand{\eeano}{\end{eqnarray*}}
\newcommand{\beq}{\begin{equation}}
\newcommand{\eeq}{\end{equation}}
\newcommand{\nonu}{\nonumber \\ }

\newcommand{\hs}[1]{\hspace{#1 mm}}

\def\cA{{\cal A}}

    \def\cN{{\cal N}}    \def\cO{{\cal O}}
        
        \def\cU{{\cal U}}

\newcommand{\BB}{{\mathbb B}}
\newcommand{\CC}{{\mathbb C}}

\newcommand{\II}{{\mathbb I}}

\newcommand{\OO}{{\mathbb O}}

\newcommand{\RR}{{\mathbb R}}
\newcommand{\TT}{{\mathbb T}}

\newcommand{\wh}[1]{\widehat{#1}}

\newcommand{\mb}[1]{\hs{4}\mbox{#1}\hs{4}}

\newcommand{\so}{\scriptscriptstyle \rm I}
\newcommand{\st}{\scriptscriptstyle \rm I\hspace{-1pt}I}

\newcommand{\la}{u}
\newcommand{\muu}{v}
\newcommand{\lac}{u^{\scriptscriptstyle C}}
\newcommand{\lab}{u^{\scriptscriptstyle B}}
\newcommand{\muc}{v^{\scriptscriptstyle C}}
\newcommand{\mub}{v^{\scriptscriptstyle B}}
\newcommand{\as}{\lambda}
\newcommand{\bla}{\bar u}
\newcommand{\bmu}{\bar v}
\newcommand{\blac}{\bar{u}^{\scriptscriptstyle C}}
\newcommand{\blab}{\bar{u}^{\scriptscriptstyle B}}
\newcommand{\bmuc}{\bar{v}^{\scriptscriptstyle C}}
\newcommand{\bmub}{\bar{v}^{\scriptscriptstyle B}}

\def\Izer{{\sf K}}

\begin{document}
\markright{ \today\dotfill \jobname\dotfill  }
\pagestyle{myheadings}
\pagestyle{plain}

\begin{center}
\textbf{\Large{Bethe vectors of gl(3)-invariant integrable models,\\[1ex]
 their scalar products and form factors}}\\
\bigskip 

{\large Eric Ragoucy\footnote{email: ragoucy@lapth.cnrs.fr}}

{\it Laboratoire de Physique Th\'eorique LAPTh, 

CNRS and Universit\'e de Savoie, BP 110, 74941 Annecy-le-Vieux Cedex, France}
\end{center}

\bigskip 

This short note corresponds to a talk given at \textit{Lie Theory and Its Applications in Physics}, (Varna, Bulgaria, June 2013) and is based on joint works with S. Belliard, S. Pakuliak and N. Slavnov, see \texttt{arXiv:1206.4931}, \texttt{arXiv:1207.0956}, \texttt{arXiv:1210.0768}, \texttt{arXiv:1211.3968} and \texttt{arXiv:1312.1488}. 
\medskip

\rightline{LAPTH-Conf-003/14
}
\section{General background}
We first expose the general algebraic framework that will be needed for our calculation. This part is not new at all, it just recasts well-known facts from QISM approach, see e.g. \cite{FadST79,BogIK93L,FadLH96} and references therein. We also use it to fix our notations.
\subsection{$R$-matrix}
As usual in integrable systems, the basic tool is the so-called $R$-matrix $R(x,y)\in V\otimes V $, where
$x,y\in\CC$ are the spectral parameters and $V=\mbox{End}(\CC^N)$ is a vector space. 
$R(x,y)$ obeys the Yang-Baxter equation, written in $V\otimes V\otimes V$:
$$
R^{{12}}(x_1,x_2)\,R^{{13}}(x_1,x_3)\,R^{{23}}(x_2,x_3)
=R^{{23}}(x_2,x_3)\,R^{{13}}(x_1,x_3)\,R^{{12}}(x_1,x_2).
$$
Here and below, we will use the auxiliary space notation: the superscripts indicate in which copies of $V$ spaces $R$ acts non trivially.
For instance, in $V\otimes V\otimes V$, we have:
$$
R^{12}(x,y)=R(x,y)\otimes\II \mb{and} R^{23}(x,y)=\II\otimes R(x,y)\,,
$$
while in $V^{\otimes N}$, we would have:
$$ 
R^{k,k+1}(x,y)=\II^{\otimes (k-1)}\otimes R(x,y) \otimes \II^{N-k-1}.
$$

\subsection{Monodromy and transfer matrices}
We define the  monodromy matrix
$$
T(x)=\sum_{i,j=1}^N{e_{ij}}\otimes T_{ij}(x)\in {\text{End}(\CC^N)}\otimes \cA[[x^{-1}]]\,,
$$
where $e_{ij}$ is the elementary $N\times N$ matrix with 1 at position $(i,j)$. $T(x)$
 obeys the commutation relations (or FRT relations)
\beq\label{eq:RTT}
R^{{12}}(x,y)\,T^{1}(x)\,T^{2}(y)=T^{2}(y)\,T^{1}(x)\,R^{{12}}(x,y).
\eeq
Through these exchange relations, the monodromy matrix generates an algebra $\cA$, defined by the choice of the $R$-matrix. 
Typically, $\cA$ is the Yangian $Y(\mathfrak{gl}_N)$ or the quantum affine group $\cU_q(\wh{\mathfrak{gl}}_N)$. The monodromy matrix
 leads to an integrable model through the transfer matrix
\beano
&&t(x)=tr_0 T^{0}(x)=\sum_{j=1}^NT_{jj}(x)\in \cA[[x^{-1}]].
\eeano
Integrability can be seen in the relation $[t(x)\,,\,t(y)]=0$, that is valid at the algebraic level (i.e. in the $\cA$
algebra), due to the relations \eqref{eq:RTT}. 

In the following, we will deal with the Yangian $Y(\mathfrak{gl}_3)$, based on the $SU(3)$-invariant $R$-matrix
$$
 R(x,y)=\mathbf{I}+g(x,y)\mathbf{P}\in End(\CC^3)\otimes End(\CC^3)
 \quad\mbox{and}\quad g(x,y)=\frac{c}{x-y},
$$
where
 $\mathbf{I}$ is the identity matrix, $\mathbf{P}$ is the permutation matrix between two spaces $End(\CC^3)$, 
 and $c$ is a constant. Note however that many properties will be also valid for the trigonometric $R$-matrix associated to the quantum group $\cU_q(\wh{\mathfrak{gl}}_3)$, and also for $Y(\mathfrak{gl}_N)$ or $\cU_q(\wh{\mathfrak{gl}}_N)$ algebras, see below.

 \subsection{Choice of a physical model}
 The choice of a representation for the algebra $\cA$ leads to a physical model. For instance, 
 taking for the monodromy and transfer matrices, the usual form
 $$
t(x)=tr_{ 0} T^{0}(x)=tr_{ 0} R^{01}(x,0)\,R^{02}(x,0)\,\cdots R^{0L}(x,0)\in (\mbox{End}(\CC^N))^{\otimes L}\,,\quad 
$$
we get an Hamiltonian acting on $L$ copies of the fundamental representation of $\cA$, ${(\CC^N)^{\otimes L}}$: it is the generalized $\mathfrak{gl}_N$-XXX or $\mathfrak{gl}_N$-XXZ closed spin chain with $L$ sites.

To summarize this algebraic part, we have a two step procedure for the determination of a physical model:
\begin{itemize}
\item The choice of an $R$-matrix, that fixes the algebra we are dealing with, that is to say the interaction in the bulk of the spin chain (leading to XXX, XXZ, ... models);
 \item
The choice of the "spin content" of the chain, that is given by the choice of the representations of the algebra, in our context the form of the monodromy matrix.
\end{itemize}

Here, as already stated, we will deal with $\cA=Y(\mathfrak{gl}_3)$. However, to be as general (and algebraic) as possible, we will not fix the representation we act on, and just assume that it is highest weight:
$$ 
T_{jj}(w)|0\rangle=\as_j(w)|0\rangle,\ j=1,2,3 \qquad   T_{ij}(w)|0\rangle=0,\qquad 1\leq i<j\leq3
$$
for some arbitrary series $\as_j(w)$, $j=1,2,3$.
Up to a rescaling $T(w)\to \as_2^{-1}(w)T(w)$, we will only need the ratios
$$
 r_1(w)=\frac{\as_1(w)}{\as_2(w)}, \qquad  r_3(w)=\frac{\as_3(w)}{\as_2(w)}.
$$
where $r_1$ and $r_3$ are free functional parameters.

\subsection{Aim}
The purpose in integrable systems is twofold:
\begin{enumerate}
\item Compute the Bethe vectors (BVs), eigenvectors of $t(x)$
$$
t(x)\,\BB^{a,b}(\bar u,\bar v) = \tau(x|\bar u,\bar v)\,\BB^{a,b}(\bar u,\bar v).
$$
This part is well-understood and is done using the algebraic Bethe ansatz method. It leads to the celebrated 
Bethe ansatz eqs (BAE).
\item\label{correla} Compute correlation functions $< \cO_1\cdots\cO_n>$ for some local operators $\cO_j$. This calculation can be decomposed in four steps:
\begin{enumerate}
\item Express the operators $\cO_j$ in terms of monodromy entries $T_{kl}(x)$;
\item Action of $T_{ij}(\bar x)$ on $\BB^{a,b}(\bar u,\bar v) $;
\item Scalar product of off-shell BVs (without BAE);
\item Form factors $\CC^{a,b}(\bar t,\bar s)T_{ij}(\bar x)\BB^{a,b}(\bar u,\bar v)$.
\end{enumerate}
\end{enumerate}
In part \ref{correla}, one needs to find \textsl{simple} (i.e. factorized) expressions in order to be able to take teh thermodynamical limit and extract the asymptotic behavior of the correlation functions.

Here, we will present these two parts for the model based on $Y(\mathfrak{gl}_3)$.
The calculations are rather technical, so that we will present here the results only, and refer to the original papers for the complete calculations.
The presentation follows the plan explained above, and we will show how the techniques apply for other models in the conclusion.

\section{Notation}
Apart from the functions $g(x,y)=\frac{c}{x-y}$, $r_1(x)$ and $r_3(x)$ introduced above, we note
$$
 f(x,y)=\frac{x-y+c}{x-y}\,,\quad
 h(x,y)=\frac{f(x,y)}{g(x,y)},\quad  t(x,y)=\frac{g(x,y)}{h(x,y)}.
$$
Clearly $f(x,y)=1+g(x,y)$ but this identification is not true for the $q$-analogues of these functions, so we keep this distinction.

To make presentation lighter, we will  use the following conventions:
\begin{itemize}
\item{"bar"} always denote {sets} of variables: $\bar w$, $\bla$, $\bmu$ etc.

\item{$|.|$} is the {dimension} of a set: $\bar w=\{w_1,w_2\}\ \Rightarrow\ |\bar w|=2$, etc.

\item{Individual elements} of the sets have {latin subscripts}: $w_j$, $u_k$,  etc.

\item{Subsets} of variables are denoted by {roman indices}: $\bla_{\so}$, $\bmu_{\rm iv}$, $\bar w_{\st}$, etc.

\item{Special case:} $\bar u_j=\bar u\setminus\{u_j\}$, $\bar w_k=\bar w\setminus\{w_k\}$, etc.
\end{itemize}

We will also use shorthand notations for products of scalar functions: 
\beano
&& f(\bla_{\st},\bla_{\so})=\prod_{\la_j\in\bla_{\st}}\prod_{\la_k\in\bla_{\so}} f(\la_j,\la_k),
\quad r_1(\bla_{\st})=\prod_{\la_j\in\bla_{\st}} r_1(\la_j),\quad
 g(v_k, \bar w)= \prod_{w_j\in\bar w} g(v_k, w_j),\quad etc.
\eeano

\section{Bethe vectors}
The framework for the construction of Bethe vectors is the Nested Bethe ansatz as introduced in \cite{KulR83}. This technics is well-known, but the explicit expressions for these BVs are rather recent, so we briefly remind them here.
\subsection{On-shell Bethe vectors}
The Bethe vectors $\BB^{a,b}(\bar u;\bar v)$ depend on two sets of parameters $\bar u=\{u_1,...,u_a\}$ 
and $\bar v=\{v_1,...,v_b\}$. The superscripts $a$ and $b$ in $\BB$ indicate the cardinalities of the sets, $|\bar u|=a$ and $|\bar v|=b$. They are eigenvectors of the transfer matrix
\bea
t(x)\, \BB^{a,b}(\bar u;\bar v) &=& \tau(x|\bar u;\bar v)\, \BB^{a,b}(\bar u;\bar v),
\\
\tau(x|\bar u;\bar v) &=&  r_1(w)f(\bar u,w)+f(w,\bar u)f(\bar v,w)+r_3(w)f(w,\bar v),
\eea
provided  $\bla$ and $\bmu$ obey the {Bethe equations (BAEs)}:
\bea
r_1(\bla_{\so}) &=& \frac{f(\bla_{\so},\bla_{\st})}{f(\bla_{\st},\bla_{\so})}f(\bmu,\bla_{\so}),
\label{eq:BAE1}\\
r_3(\bmu_{\so}) &=& \frac{f(\bmu_{\st},\bmu_{\so})}{f(\bmu_{\so},\bmu_{\st})}f(\bmu_{\so},\bla)
\label{eq:BAE2}.
\eea
that hold for arbitrary partitions of the sets $\bla$ and $\bmu$ into subsets
$\{\bla_{\so},\;\bla_{\st}\}$ and $\{\bmu_{\so},\;\bmu_{\st}\}$. In that case, the BVs will be called "on-shell", while they will be called "off-shell" is the BAEs are not obeyed. Of course, in that latter case, the BVs are not eigenvector of $t(x)$.

\subsection{Dual Bethe vectors $\CC^{a,b}(\bla;\bmu)$, $|\bla|=a$, $|\bmu|=b$}
Dual BVs are constructed as left eigenvectors of the transfer matrix:
\beq  
\CC^{a,b}(\bar u;\bar v)\,t(x) = \tau(x|\bar u;\bar v)\, \CC^{a,b}(\bar u;\bar v),
\eeq
where the Bethe parameters $\bla,\bmu$ obey the BAEs \eqref{eq:BAE1}--\eqref{eq:BAE2}. Again, these dual BVs will be called on-shell when $\bla$ and $\bmu$ obey the BAEs, while they will be called off-shell dual BVs when $\bla,\bmu$ are left free.

\subsection{Trace formula}
This is a known and quite general formula, given in \cite{TV} for $\mathfrak{gl}_N$ and $\cU_q(\mathfrak{gl}_N)$ algebras, and generalized in \cite{BR1} for superalgebras. It expresses $\BB^{a,b}(\bla;\bmu)$ as a trace in $a+b$ auxiliary spaces of products of monodromy matrices:
\beq
\BB^{a,b}(\bla;\bmu) = {tr} \Big( \TT(\bla;\bmu)\,\RR(\bla;\bmu)\, {e_{21}^{\otimes a}\otimes \, e_{32}^{\otimes b}}\Big)
\in Y(\mathfrak{gl}_3),
\eeq
where $\TT$ is some product of monodromy matrices $T(x)$ and $\RR$ some product of $R$-matrices. Their explicit expression can be found in \cite{TV,BR1}.

\subsection{Recursion formulas\label{sec:recursion}}
It can be shown that the Bethe vectors also obey the following recursion relations \cite{BelPRS12b}:
\begin{eqnarray}
&&\lambda_2(u_k)f(\bar v,u_k)\BB^{{a+1,b}}(\bar u;\bar v)
\ =\ T_{12}(u_k) \BB^{{a,b}}(\bar u_k;\bar v)
+ \sum_{i=1}^b g( v_{i},u_k)f(\bar v_{i}, v_{i})  T_{13}(u_k)\BB^{{a,b-1}}(\bar u_k; \bar v_i),
\label{eq:recur-a}\qquad\\
&&\lambda_2(v_{k})f(v_{k},\bar u) \BB^{{a,b+1}}(\bar u;\bar v)=  T_{23}(v_{k}) \BB^{{a,b}}(\bar u;\bar v_{k})
+ \sum_{j=1}^a g(v_{k},u_j)f(u_{j}, \bar u_{j}) T_{13}(v_{k})\BB^{{a-1,b}}(\bar u_{j}; \bar v_{k}).
\label{eq:recur-b}
\end{eqnarray}
Let us remark that \eqref{eq:recur-a} completely determines the Bethe vectors once $\BB^{{0,b}}(\emptyset;\bar v)$ is known. In the same way, \eqref{eq:recur-b} completely determines the Bethe vectors once $\BB^{{a,0}}(\bar u;\emptyset)$
is fixed.

\subsection{Explicit formulas\label{sec:explicit}}
There is a third series of expressions for Bethe vectors, using partitions of $\bar u$ and $\bar v$ \cite{BelPRS12b}:
\bea
\BB^{a,b}(\bar u;\bar v) &=&
\sum\frac{\Izer_k(\bar v_{\so}|\bar u_{\so})}{ \as_2(\bar v_{\st})\as_2(\bar u)}
\frac{f(\bar v_{\st},\bar v_{\so})f(\bar u_{\st},\bar u_{\so})}
{f(\bar v_{\st},\bar u)f(\bar v_{\so},\bar u_{\so})}\
T_{12}(\bar u_{\st})T_{13}(\bar u_{\so}) T_{23}(\bar v_{\st})\,
|0\rangle\,,
\\
\BB^{a,b}(\bar u;\bar v) &=& \sum
\frac{\Izer_k(\bar v_{\so}| \bar u_{\so})}{ \as_2(\bar u_{\st})\as_2(\bar v)}
\frac{ f(\bar v_{\so},\bar v_{\st})f(\bar u_{\so},\bar u_{\st})}{f(\bar v_{\so},\bar u_{\so})f(\bar v,\bar u_{\st})}\ 
T_{23}(\bar v_{\st})T_{13}(\bar v_{\so}) T_{12}(\bar u_{\st})\,
|0\rangle\,,
\\
\BB^{a,b}(\bla;\bmu) &=& \sum \frac{\Izer_{k}(\bmu_{\so}|\bla_{\so})}{\as_2(\bar v_{\st})\as_2(\bar u)}
\frac{f(\bmu_{\st},\bmu_{\so})f(\bla_{\so},\bla_{\st})}{f(\bmu,\bla)}\
T_{13}(\bla_{\so})T_{12}(\bla_{\st})T_{23}(\bmu_{\st})\,|0\rangle\,,
\\
\BB^{a,b}(\bla;\bmu) &=& \sum \frac{\Izer_{k}({\bmu}_{\so}|{\bla}_{\so})}{\as_2(\bar u_{\st})\as_2(\bar v)}
\frac{f(\bmu_{\st},\bmu_{\so})f(\bla_{\so},\bla_{\st})}{f(\bmu,\bla)}\
T_{13}({\bmu}_{\so})T_{23}({\bmu}_{\st})T_{12}({\bla_{\st}})\,|0\rangle.
\eea

The sums are taken over partitions of the sets 
$\bar u\Rightarrow\{\bar u_{\so},\bar u_{\st}\}$ and $\bar v\Rightarrow\{\bar v_{\so}$, $\bar v_{\st}\}$ with the condition
$0\leq |\bar u_{\so}|=|\bar v_{\so}|=k\leq\mbox{min}(a,b)$. 

$ \Izer_k(\bar v_{\so}|\bar u_{\so})$ is the Izergin--Korepin determinant \cite{Ize87}
\beq
 \Izer_k(\bar x|\bar y)=\prod_{\ell<m}^k g(x_\ell,x_m)g(y_m,y_\ell)\cdot h(\bar x,\bar y)
 \; \det_k\left[ t(x_i,y_j) \right].
\eeq

\subsection{All these formulas are related}
Let us stress that all the above formulas define the same Bethe vectors, should they be on-shell or off-shell. For instance, one can show that
\begin{itemize}
\item The explicit expressions obey the recursion formulas;
\item The trace formula obeys the recursion formulas too;
\item Recursion formulas can be obtained starting from the trace formula.
\end{itemize}
Depending on the calculation, one can then freely choose any of these expression to prove a formula or a property of BVs.

\section{Correlation functions}
We now turn to the second step of our program, that is, for a local operator $\cO$, how to compute its mean value? As
a first step, we are led with the following question:\\

\centerline{\textbf{How to compute $\cO_{\CC,\BB}=\langle\CC|\cO|\BB\rangle$?}}
\medskip

Assuming that $\{|\BB\rangle\}$ forms a complete basis (of transfer matrix eigenspaces), we have
\beq\label{eq:Opsi}
\cO|\BB\rangle=\sum_{\BB'} \OO_{\BB\BB'}|\BB'\rangle,
\eeq
so that we "only" need $\langle\CC|\BB'\rangle$ and of course the decomposition \eqref{eq:Opsi}.

Now, for a spin chain of length $L$ and based on $\mathfrak{gl}_N$-fundamental representations, local operators have a decomposition\footnote{The same ideas can be applied for a general spin chain, using an adapted basis.}
\beq
\cO=\sum_{\ell=1}^L\sum_{i,j=1}^N \cO^{(\ell)}_{ij}\, e_{ij}^{\ell},
\label{eq:O-decomp}
\eeq
where $e_{ij}^{\ell}$ is the elementary matrix $e_{ij}$ at site $\ell$.
Then, everything boils down to the calculation of $\langle\CC|e_{ij}^{\ell}|\BB\rangle$.
\medskip

A further simplification occurs because of QISM. Indeed, the
expression of $e_{ij}^{\ell}$, $i,j=1,2,...,N$ and $\ell=1,...L$ is known in terms of monodromy entries $T_{kl}(x)$, 
$k,l=1,...,N$ \cite{MaiT00}:
\beq
 e^{\ell}_{ij} =(t(0))^{\ell-1} \, T_{ij}(0)\,(t(0))^{-\ell}.
\label{eq:QISM}
\eeq
Then, from \eqref{eq:O-decomp} and \eqref{eq:QISM}, if we can compute $T_{kl}(x)\BB^{a,b}(\bla;\bmu)$ and $\CC^{a,b}(\bar w;\bar z)\BB^{a,b}(\bar u;\bar v)$, we are able to compute any correlation function. The following sections are devoted to the calculation of these two fundamental quantities in the case $N=3$.

\section{Multiple actions of $T_{ij}(\bar x)$ on $\mathbb{B}^{a,b}(\bla;\bmu)$}
Using the explicit expressions of section \ref{sec:explicit}, we were able in \cite{BelPRS12c} to compute explicitly the actions of $T_{ij}(\bar x)$ on $\mathbb{B}^{a,b}(\bla;\bmu)$. Denoting $\{\bla,\bar x\}=\bar\eta$, $\{\bmu,\bar x\}=\bar\xi$ and  the cardinalities by
$|\bar x|=n$, $|\bar\eta|=a+n$
and $|\bar\xi|=b+n$, we have 
\bea
 T_{13}(\bar x)\mathbb{B}^{a,b}(\bla;\bmu) &=& \lambda_2(\bar x)\,\mathbb{B}^{a+n,b+n}(\bar\eta;\bar\xi),
\label{actionT13}
\\
\label{actionT12}
 T_{12}(\bar x)\mathbb{B}^{a,b}(\bla;\bmu) &=& (-1)^n\lambda_2(\bar x)\,\sum
 f(\bar\xi_{\st},\bar\xi_{\so})\Izer_n(\bar\xi_{\so}|\bar x+c)\,
 \mathbb{B}^{a+n,b}(\bar\eta;\bar\xi_{\st}),
\\
\label{actionT23}
 T_{23}(\bar x)\mathbb{B}^{a,b}(\bla;\bmu) &=& (-1)^n\lambda_2(\bar x)\,\sum
 f(\bar\eta_{\so},\bar\eta_{\st})\Izer_n(\bar x|\bar\eta_{\so}+c)\,
 \mathbb{B}^{a,b+n}(\bar\eta_{\st};\bar\xi).
\eea
In \eqref{actionT12}, the sum is on partitions $\bar\xi=\{\bar\xi_{\so};\bar\xi_{\st}\}$ with $|\bar\xi_{\so}|=n$, while in \eqref{actionT23}, 
the sum is on partitions $\bar\eta=\{\bar\eta_{\so};\bar\eta_{\st}\}$ with $|\bar\eta_{\so}|=n$.
Similar expressions can be obtained for any $T_{ij}(\bar x)$ and for dual BVs, see \cite{BelPRS12c}.

Remark that the relations \eqref{actionT12} and \eqref{actionT23} imply recursion relations of section \ref{sec:recursion} as a subcase (for $n=1$).

Since the action of $T_{ij}(\bar x)$ operators on BVs gives back BVs (that are a priori off-shell), it remains to compute 
scalar products of BVs to get the full form factor expression.

\section{Scalar products of BVs}
In this section, we provide expression for the scalar product
\beq
 \mathcal{S}_{a,b}\equiv\mathcal{S}_{a,b}(\blac,\blab|\bmuc,\bmub)=
\CC^{a,b}(\blac;\bmuc)\,\BB^{a,b}(\blab;\bmub),
\eeq
where $\CC^{a,b}(\blac;\bmuc)$ and $\BB^{a,b}(\blab;\bmub)$ are general (dual) BVs.
Let us stress that the superscripts $^B$ and $^C$ are used to denote \textit{different} sets of (Bethe) parameters, completely independent one from each other.

\subsection{Reshetikhin's formula}
There is a well-known formula, due to Reshetikhin \cite{Res86}, and valid for $\mathfrak{gl}_N$:
\begin{align}
\mathcal{S}_{a,b}=&\sum r_1(\blab_{\so})r_1(\blac_{\st})r_3(\bmub_{\so}) r_3(\bmuc_{\st})
f(\blac_{\so},\blac_{\st})  f(\blab_{\st},\blab_{\so})  f(\bmuc_{\st},\bmuc_{\so})
f(\bmub_{\so},\bmub_{\st})
f(\bmuc_{\so},\blac_{\so})f(\bmub_{\st},\blab_{\st}) \nonumber\\
&\qquad\times Z_{a-k,n}(\blac_{\st};\blab_{\st}|\bmuc_{\so};\bmub_{\so})
Z_{k,b-n}(\blab_{\so};\blac_{\so}|\bmub_{\st};\bmuc_{\st}),
\label{eq:resh}
\end{align}
where the sum is on partitions $\blab=\{\blab_{\so},\blab_{\st}\}$, $\blac=\{\blac_{\so},\blac_{\st}\}$ with $|\blab_{\so}|=|\blac_{\so}|=k$ for  $k=0,\dots,a$ 
$\bmub=\{\bmub_{\so},\bmub_{\st}\}$, $\bmuc=\{\bmuc_{\so},\bmuc_{\st}\}$ with $|\bmub_{\so}|=|\bmuc_{\so}|=n$ for $n=0,\dots,b$.

$Z_{a,b}$ are the so-called highest coefficients
\beq\label{eq:highest}
  Z_{a,b}(\bar t;\bar x|\bar s; \bar y)=(-1)^b\sum
 K_b(\bar s-c|\bar w_{\so})K_a(\bar w_{\st}|\bar t)
  K_b(\bar y|\bar w_{\so})f(\bar w_{\so},\bar w_{\st}),
\eeq
where the sum is done over partitions of $\bar w=\{\bar s,\bar x\}$ into
subsets $\bar w_{\so}$ and $\bar w_{\st}$ with $|\bar w_{\so}|=b$.

The formula is valid for a general scalar product, but as it stands, $\mathcal{S}_{a,b}$ is difficult to handle. To compute e.g. the thermodynamical limit of such formula, and to use it for the calculation of correlation functions, one needs to find a factorized form, containing only one determinant. It was done for the $\mathfrak{gl}_2$ case \cite{IzKo}, but for $\mathfrak{gl}_3$ (and a fortiori for $\mathfrak{gl}_N$) no such formula is known yet. 
However, in some particular cases, there exists such a formula: 
\begin{enumerate}
\item When computing the norm of a Bethe vector that is assumed to be on-shell, such an expression was obtained by Reshetikhin \cite{Res86};
\item A nice factorized expression was obtained in \cite{Whe12}, when some of the Bethe parameters tend to infinity; 
\item \label{item:scal} When the BV is on-shell and the dual BV is "twisted on-shell" (see below), we were able to to get a simplified expression \cite{BelPRS12b};
\item \label{item:highest} In \cite{BelPRS12a}, we provided  different expressions for the highest coefficients \eqref{eq:highest}; 
\item An interesting multiple integral expression for the scalar product of an on-shell and an off-shell BV  was recently obtained in \cite{Whe13}.
\end{enumerate}
We present the points \ref{item:scal}  and \ref{item:highest} in the two following sections.

\subsection{Highest coefficients}
Highest coefficients were introduced by Reshetekhin \cite{Res86} and play a central role in the expression of the scalar product of Bethe vectors. In fact, they can be viewed as partition functions of a statistical models with some particular boundary conditions. It is thus important to get different forms for them. We give here some examples of such formulas, a more complete list can be found in \cite{BelPRS12a}.
\paragraph{Sums on partitions.} There are different series of expressions for the highest coefficients. A first series is given by sums over partitions. The expression \eqref{eq:highest} is a first example of such formulas. Another example is given by
\beq
Z_{a,b}(\bar t;\bar x|\bar s;\bar y)=(-1)^af(\bar y,\bar x)f(\bar s,\bar t)
\sum K_a(\bar t-c|\bar\eta_{\so})K_a(\bar x|\bar\eta_{\so})K_b(\bar\eta_{\st}-c|\bar s)f(\bar\eta_{\so},\bar\eta_{\st}),
\eeq
where $\bar\eta=\{\bar y+c,\;\bar t\}$. The sum is taken with respect to partitions of the set $\bar\eta$ into
subsets $\bar\eta_{\so}$ and $\bar\eta_{\st}$ with $\#\bar\eta_{\so}=a$. 

\paragraph{Recursion formulas.} 
The most important property of the highest coefficient $Z_{a,b}$ is that its residues in its poles can be expressed in terms of $Z_{a-1,b}$ or $Z_{a,b-1}$. Since $Z_{a,b}$ is a rational function in all its variables, this property allows us to fix it  unambiguously, provided we know  some initial condition.  It is easy to see that for $a=0$ or $b=0$  $Z_{a,b}$ coincides with $K_n$:
\beq
Z_{a,0}(\bar t;\bar x|\emptyset;\emptyset)=K_{a}(\bar x|\bar t),\qquad
Z_{0,b}(\emptyset;\emptyset|\bar s;\bar y)=K_b(\bar y|\bar s).
\eeq

Consider $Z_{a,b}$ as a function of $s_b$ with all other variables fixed.
Then it has simple poles at $s_b= y_m$, $m=1,\dots,b$ and  $s_b= t_\ell$, $\ell=1,\dots,a$. Due to the
symmetry of $Z_{a,b}$ over $\bar y$ and over $\bar t$ it is enough to
find the residues at $s_b=y_b$ and $s_b=t_a$. These residues are given by:
\bea
\Bigl.\mbox{Res}\, Z_{a,b}(\bar t;\bar x|\bar s;\bar y)\Bigr|_{ s_b= y_b} &=&
-cf( y_b,\bar s_{b})f(\bar y_{b}, y_b)f( y_b,\bar x)
Z_{a,b-1}(\bar t;\bar x|\bar s_{b};\bar y_{b}),
\\
\Bigl.\mbox{Res}\, Z_{a,b}(\bar t;\bar x|\bar s;\bar y)\Bigr|_{ s_b= t_a} &=&
c f( \bar s_{b}, t_a)f( t_a, \bar t_{a})\sum_{p=1}^a g(x_p, t_a)f(\bar x_{p},x_p)
Z_{a-1,b}(\bar t_{a};\bar x_{p}|\{\bar s_{b},\; x_p\};\bar y_{b}),
\eea
where $\bar s_{b}=\bar s\setminus s_{b}$,  $\bar y_{b}=\bar y\setminus y_{b}$, etc.

\paragraph{Contour integral.} 
There exists several representations for $Z_{a,b}$ in terms
of multiple contour integrals of Cauchy type. Here, we give only one possible integral as example:
\beq\label{eq:contour}
Z_{a,b}(\bar t;\bar x|\bar s;\bar y) = \frac1{(2\pi ic)^bb!} \oint\limits_{\bar w} K_b(\bar s-c|\bar z)
K_b(\bar y|\bar z) K_{a+b}(\bar w|\bar t,\bar z+c) f(\bar z,\bar w) \mathcal{F}_b(\bar z) \,d\bar z,
\eeq
where we have a $b$-fold integral and
$$\mathcal{F}_b(\bar z)=\prod_{j=1}^b f^{-1}(z_j,\bar z_j).$$
Other expressions of the type \eqref{eq:contour}, or implying $a$-fold integrals can be found in \cite{BelPRS12a}.

\subsection{Scalar product for twisted Bethe vectors}
Here we consider an on-shell Bethe vector, eigenvector of the transfer matrix
\beq
t(x)\, \BB^{a,b}(\bar u^B;\bar v^B) = \tau(x|\bar u^B,\bar v^B)\, \BB^{a,b}(\bar u^B;\bar v^B),
\eeq
where the Bethe parameters $\{\bar u^B;\bar v^B\}$ obey the Bethe equations \eqref{eq:BAE1}-\eqref{eq:BAE2}.
We also introduce, for any complex number $\kappa$, a twisted transfer matrix
\beq
t_{{\kappa}}(x) = T_{11}(x)+{\kappa}T_{22}(x)+T_{33}(x)=tr\big({M}\,T(x)\big)\mb{with} 
{M}=\left(\begin{array}{ccc} 1 & 0 &0 \\ 0& {\kappa} & 0 \\ 0& 0& 1\end{array}\right)
\eeq
 and its \textit{twisted} dual on-shell Bethe vector 
 \beq
\CC_{{\kappa}}^{a,b}(\bar u^C;\bar v^C)\,t_{{\kappa}}(x)  =
\tau_{{\kappa}}(x|\bar u^C,\bar v^C)\, \CC_{{\kappa}}^{a,b}(\bar u^C;\bar v^C).
\eeq
It is an eigenvector of $t_{{\kappa}}(x)$
when the Bethe parameters $\bar u^C$, $\bar v^C$ obey the \textit{twisted} BAEs
\bea
r_1(\bla_{\so}) &=& \kappa\,\frac{f(\bla_{\so},\bla_{\st})}{f(\bla_{\st},\bla_{\so})}f(\bmu,\bla_{\so}),
\label{eq:BAEtw1}\\
r_3(\bmu_{\so}) &=& \kappa\,\frac{f(\bmu_{\st},\bmu_{\so})}{f(\bmu_{\so},\bmu_{\st})}f(\bmu_{\so},\bla)
\label{eq:BAEtw2}.
\eea
Let us stress that the superscripts $^B$ and $^C$ are there to distinguish the Bethe parameters of $\BB^{ab}$ 
from those of $\CC^{ab}$. In other words, the Bethe parameters $\{\bar u^B, \bar v^B\}$ are a priori not related to 
$\{\bar u^C, \bar v^C\}$.

In \cite{BelPRS12b}, we obtained an expression for the scalar product
\beq
{ \mathcal{S}_{a,b}\equiv\mathcal{S}_{a,b}(\blac,\blab|\bmuc,\bmub)=
\CC_{{\kappa}}^{a,b}(\blac;\bmuc)\,\BB^{a,b}(\blab;\bmub)}.
\eeq

Indeed, the scalar product can be written as
\bea
\mathcal{S}_{a,b}&=& f(\bmuc,\blac)f(\bmub,\blab)t(\bmuc,\blab)
 \Delta'_a(\blac)\Delta_a(\blab)\Delta'_b(\bmuc)\Delta_b(\bmub)\, \det_{a+b}\mathcal{N},
\eea
where 
$$
\Delta'_n(\bar x)
=\prod_{j>k}^n g(x_j,x_k),\qquad {\Delta}_n(\bar y)=\prod_{j<k}^n g(y_j,y_k).
$$
and $\mathcal{N}$ is a block-matrix of the size $(a+b)\times(a+b)$,
$$
 \mathcal{N}=\left(\begin{array}{cc}
 {\displaystyle \mathcal{N}^{(\la)}(\lac_j,\lab_k)}&
 {\displaystyle \mathcal{N}^{(\la)}(\lac_j,\muc_k)}\\
 {\displaystyle \mathcal{N}^{(\muu)}(\mub_j,\lab_k)}&
 {\displaystyle \mathcal{N}^{(\muu)}(\mub_j,\muc_k)}
 \end{array}\right)=\left(\begin{array}{c|c}
a\times a & a\times b\\
\hline
b\times a & b\times b
 \end{array}\right),
$$
whose full expression is given in appendix \ref{app:N}. We show below  how this expression can give rise to a factorized expression for form factors of the model.

\subsubsection*{Expression for a general twist}
A similar expression for $\mathcal{S}_{a,b}$ can be obtained when considering a general twist
${\bar\kappa}=(\kappa_1,\kappa_2,\kappa_3)$ of the transfer matrix
$$
t_{{\bar\kappa}}(x)={\kappa_1}T_{11}(x)+{\kappa_2}T_{22}(x)+{\kappa_3}T_{33}(x).
$$
However, in that case, the expression is valid only up to terms $(\kappa_i-1)(\kappa_j-1)$, $i,j=1,2,3$, that are irrelevant for our purpose, as we shall see below.
For further application it is useful to write the system of twisted Bethe equations in the logarithmic form. Let us define
\beq\label{Phi-1}
\Phi_j=\log r_1(\lac_{j})-\log \left(\frac{f(\lac_{j},\blac_{j})}{f(\blac_{j},\lac_{j})}\right) -\log f(\bmuc,\lac_{j}),
\qquad j=1,\dots,a,
\eeq
and
\beq\label{Phi-2}
\Phi_{j+a}=\log r_3(\muc_{j})-\log \left(\frac{f(\bmuc_{j},\muc_{j})}{f(\muc_{j},\bmuc_{j})}\right)-\log f(\muc_{j},\blac),
\qquad j=1,\dots,b.
\eeq
Then the system of twisted Bethe equations for general ${\bar\kappa}$ takes the form
\beq\label{Log-TBE}
\begin{array}{l}
\Phi_j=\log\kappa_2-\log\kappa_1+2\pi i \ell_j,\qquad j=1,\dots,a,
\\
\Phi_{j+a}=\log\kappa_2-\log\kappa_3+2\pi i m_j,\qquad j=1,\dots,b,
\end{array}
\eeq
where $\ell_j$ and $m_j$ are some integers. 

\section{Form factors}
We present now the calculation \cite{BelPRS13a} the form factor of the diagonal elements $T_{ss}(z)$
\beq\label{eq:formfactor}
 \mathcal{F}_{a,b}^{(s)}(z)\equiv\mathcal{F}_{a,b}^{(s)}(z|\blac,\bmuc;\blab,\bmub)=
 \mathbb{C}^{a,b}(\blac;\bmuc)T_{ss}(z)\mathbb{B}^{a,b}(\blab;\bmub),
\eeq
where both $\mathbb{C}^{a,b}(\blac;\bmuc)$ and $\mathbb{B}^{a,b}(\blab;\bmub)$ are on-shell
Bethe vectors. Form factors for off-diagonal elements $T_{j,j+1}(z)$ and $T_{j+1,j}(z)$ have been given in \cite{PRS13b}. The form factors associated to $T_{13}(z)$ and $T_{31}(z)$ remain to be done. Of course, the ultimate goal would be to find a simple expression for the form factor when the Bethe vector $\mathbb{B}^{a,b}(\blab;\bmub)$ and/or the dual Bethe vector $\mathbb{C}^{a,b}(\blac;\bmuc)$ are off-shell. Up to now, such an expression is still missing.  

A priori, from the knowledge of the actions \eqref{actionT13}-\eqref{actionT23} and the scalar products \eqref{eq:resh}, 
we can deduce an expression of the form factor. However, the expression is rather complicated and difficult to handle. Fortunately, one can get another simpler form using the following trick. 

Let us consider $\mathbb{C}_{\bar\kappa}^{a,b}(\blac;\bmuc)$ a twisted on-shell Bethe
vector such that 
\beq
\mathbb{C}_{\bar\kappa}^{a,b}(\blac;\bmuc)|_{\bar\kappa=1}=\mathbb{C}^{a,b}(\blac;\bmuc).
\eeq
Then, the form factor \eqref{eq:formfactor} can be expresssed as
$$
\mathcal{F}^{(s)}(z|\blac,\bmuc;\blab,\bmub)=
\frac{d Q_{\bar\kappa}(z)}{d\kappa_s}\Bigl.\Bigr|_{\bar\kappa=1}\,,\qquad s=1,2,3
$$
where ${\bar\kappa=1}$ means  $\kappa_1=\kappa_2=\kappa_3=1$ and
\beano
Q_{\bar\kappa}(z)&=&\mathbb{C}_{\bar\kappa}^{a,b}(\blac;\bmuc) \bigl(t_{\bar\kappa}(z)-t(z)\bigr)\mathbb{B}^{a,b}(\blab;\bmub)
\\
&=&  \bigl(\tau_{\kappa}(z|\bar u^C,\bar v^C)-\tau(z|\bar u^B,\bar v^B)\bigr)\mathbb{C}_{\bar\kappa}^{a,b}(\blac;\bmuc)\mathbb{B}^{a,b}(\blab;\bmub).
\eeano
Then, it is clear that all depends on the expression of the scalar product $\mathbb{C}_{\bar\kappa}^{a,b}(\blac;\bmuc)\mathbb{B}^{a,b}(\blab;\bmub)$, and that we need to know this scalar product only up to terms $(\kappa_i-1)(\kappa_j-1)$, $i,j=1,2,3$.
Depending on whether $\mathbb{C}^{a,b}(\blac;\bmuc)$ is $\left(\mathbb{B}^{a,b}(\blab;\bmub)\right)^\dag$ or not, we get two different expressions:
\paragraph{When $\mathbb{C}^{a,b}(\blac;\bmuc)=\left(\mathbb{B}^{a,b}(\blab;\bmub)\right)^\dag$}
\bea
\mathcal{F}^{(s)}(z|\bla,\bmu;\bla,\bmu) &=&
\|\mathbb{B}^{a,b}(\bla;\bmu)\|^2\;\frac{d \tau_{\bar\kappa}(z|\blac;\bmuc)}{d\kappa_s}\Big|_{\bar\kappa=1}
\nonu
&=& (-1)^ac^{a+b}f(\bmu,\bla)\prod_{j=1}^a f(\la_j,\bar\la_j)
 \prod_{k=1}^b f(\muu_k,\bar\muu_k)\ \det_{a+b+1}\Theta^{(s)}(z),
\eea
where $\Theta^{(s)}(z)$ is an $(a+b+1)\times(a+b+1)$ matrix given in appendix \ref{app:Theta}.
\paragraph{When $\mathbb{C}^{a,b}(\blac;\bmuc)\neq\left(\mathbb{B}^{a,b}(\blab;\bmub)\right)^\dag$}
\beano
\mathcal{F}_{a,b}^{(s)}(z|\blac,\bmuc;\blab,\bmub)&=&\Big((\tau(z|\blac;\bmuc)-\tau(z|\blab;\bmub)\Big)\ 
\frac{d}{d\kappa_s}\,
\left(\mathbb{C}_{\bar\kappa}^{a,b}(\blac;\bmuc)\mathbb{B}^{a,b}(\blab;\bmub)\right)\,\Big|_{\bar\kappa=1}
\nonu
&=&
\frac{\tau(z|\blac;\bmuc)-\tau(z|\blab;\bmub)}{\Omega_{p}}\ t(\bmuc,\blab)
 \Delta'_a(\blac)\Delta_a(\blab)\Delta'_b(\bmuc)\Delta_b(\bmub)
\nonu
&&\times \det_{a+b}\mathcal{N}^{(s,p)},
\eeano
The integer $p$ is such that $\Omega_p\neq0$ where
\beq
\begin{array}{rcll}
{\displaystyle \Omega_k} &=& {\displaystyle \prod\limits_{\ell=1}^a(\lac_k-\lab_\ell)
\prod\limits_{\ell=1\atop{\ell\ne k}}^a(\lac_k-\lac_\ell)^{-1},}
&{\displaystyle \qquad k=1,\dots,a,}\\
{\displaystyle \Omega_{a+k}} &=& {\displaystyle \prod\limits_{m=1}^b(\mub_k-\muc_m)
\prod\limits_{m=1\atop{m\ne k}}^b(\mub_k-\mub_m)^{-1},}
&{\displaystyle \qquad k=1,\dots,b.}
\end{array}
\eeq
The matrix $\mathcal{N}^{(s,p)}$ has a special $p^{th}$ row,
but its determinant is independent of $p$. The form of $\mathcal{N}^{(s,p)}$ is given in appendix \ref{app:Nsp}.


\section{Conclusion}
For models with a $\mathfrak{gl}_3$ invariant $R$-matrix, we have presented several
explicit expressions for (off-shell) Bethe vectors and duals BVs. We have also computed the
multiple action of monodromy elements on these BVs. 
Both results are presented in term of Izergin-Korepin determinants and 
sums over partitions of sets of Bethe parameters.

In a second step, we calculated the scalar product of (twisted) on-shell BVs and the
 form factors of $T_{ss}(x)$, $s=1,2,3$, of $T_{j,j+1}(x)$ and of $T_{j+1,j}(x)$, $j=1,2$.
Both results were given in term of a single determinant (and product of scalar functions).
\medskip

The ultimate goal is to obtain a single determinant expression for the correlation functions of the model, so as to study the thermodynamical limit and their asymptotics. Of course, to get to that point a lot remains to be done. For instance, it remains to compute the form factors of $T_{13}(x)$ and $T_{31}(x)$. The 
calculation of the scalar product of generic off-shell BVs (as a single determinant) is also lacking.
\medskip

 Certainly, a generalization to other integrable models is wanted. As a first step, we started to investigate the case of $\mathfrak{gl}_3$ XXZ spin chain (i.e. based on the $R$-matrix of $\cU_q(\mathfrak{gl}_3)$):
\begin{enumerate}
\item The multiple action of $T_{ij}(x)$ generators on BVs was performed in \cite{BelPRS13d};
\item The calculation of the highest coefficient was done in \cite{PRS13c};
\item A Reshetikhin-like formula for scalar products of the $\cU_q(\mathfrak{gl}_3)$ model is given in \cite{PRS14a}.
\end{enumerate}

Let us remark that to obtain these results, we used the current realization of $\cU_q(\mathfrak{gl}_3)$ and the construction 
of Khoroshkin, Pakuliak and collaborators for BVs in this presentation \cite{KhEOPS}. 
This construction is valid for $\cU_q(\mathfrak{gl}_N)$:
a link between the current presentation of BVs and the 
explicit expression of BVs using the monodromy matrix for $\cU_q(\mathfrak{gl}_N)$ is done in \cite{PRS13e}. The use of a morphism between $\cU_q(\mathfrak{gl}_N)$ and $\cU_{q^{-1}}(\mathfrak{gl}_N)$ is essential in this construction.

\appendix
\section{The matrix $\cN$\label{app:N}}
\paragraph{Diagonal blocks}
\begin{multline*}
\mathcal{N}^{(\la)}(\lac_j,\lab_k) = h(\bmuc,\lab_k)h(\lab_k,\blac)
\Big[\kappa t(\lab_k,\lac_j)\\
+t(\lac_j,\lab_k)\frac{f(\bmub,\lab_k)}{f(\bmuc,\lab_k)}
\frac{h(\blac,\lab_k)h(\lab_k,\blab)}{h(\lab_k,\blac)h(\blab,\lab_k)} \Big]
\quad\mb{{$a\times a$ block}}
\end{multline*}

\begin{multline*}
\mathcal{N}^{(\muu)}(\mub_j,\muc_k) = h(\muc_k,\blab)h(\bmub,\muc_k)\Big[t(\mub_j,\muc_k)\\
+
\kappa t(\muc_k,\mub_j)\frac{f(\muc_k,\blac)}{f(\muc_k,\blab)}
\frac{h(\bmuc,\muc_k )h(\muc_k,\bmub)}{h(\muc_k,\bmuc)h(\bmub,\muc_k )}  \Big]
\quad\mb{{$b\times b$ block}}
\end{multline*}

\paragraph{Off-diagonal blocks}
\begin{align*}
 \mathcal{N}^{(\la)}(\lac_j,\muc_k)&=\kappa
  t(\muc_k,\lac_j)h(\muc_k,\blac)h(\bmuc,\muc_k)
\quad\mb{{ \ $a\times b$ block}}
\\
 \mathcal{N}^{(\muu)}(\mub_j,\lab_k)&=
 t(\mub_j,\lab_k)h(\bmub,\lab_k )h(\lab_k,\blab)
\qquad\mb{{$b\times a$ block}}
\end{align*}

\section{The matrix $\Theta^{(s)}$\label{app:Theta}}
First of all we define an  $(a+b)\times(a+b)$ matrix $\theta$ with the entries
\beq\label{theta}
\theta_{j,k}=\left.\frac{\partial\Phi_j}{\partial \lac_k}\right|_{\blac=\bla\atop{\bmuc=\bmu}},\qquad k=1,\dots,a;\quad\text{and}\quad
\theta_{j,k+a}=\left.\frac{\partial\Phi_j}{\partial \muc_k}\right|_{\blac=\bla\atop{\bmuc=\bmu}},\qquad k=1,\dots,b,
\eeq
where the $\Phi_j$ are given by  \eqref{Phi-1} and \eqref{Phi-2}.

Then we extend the matrix $\theta$ to an $(a+b+1)\times(a+b+1)$ matrix $\Theta^{(s)}$ with $s=1,2,3$, by adding one
row and one column
\beq\label{Theta}
\begin{array}{llll}
\displaystyle \Theta^{(s)}_{j,k}=\theta_{j,k}, \quad\qquad\qquad\qquad j,k=1,\dots,a+b,&
\nonumber\\[2.1ex]
\displaystyle\Theta^{(s)}_{a+b+1,k}=\frac{\partial\tau(z|\bla,\bmu)}{\partial\la_k},\qquad k=1,\dots, a,\qquad
&\displaystyle\Theta^{(s)}_{a+b+1,a+k}=\frac{\partial\tau(z|\bla,\bmu)}{\partial\muu_k},\qquad k=1,\dots, b,
\nonumber\\[2.1ex]
\displaystyle\Theta^{(s)}_{j,a+b+1}=\delta_{s1}-\delta_{s2}\qquad\quad j=1,\dots, a,\qquad
&\displaystyle\Theta^{(s)}_{j+a,a+b+1}=\delta_{s3}-\delta_{s2}\qquad\quad j=1,\dots, b,
\nonumber\\[2.1ex]
\displaystyle\Theta^{(s)}_{a+b+1,a+b+1}=\left.\frac{\partial\tau_{\bar\kappa}(z|\blac,\bmuc)}{\partial\kappa_s}
\right|_{\blac=\bla\atop{\bmuc=\bmu}}.&
\end{array}
\eeq
Here the $\delta_{sk}$ are  Kronecker deltas. Notice that $\Theta^{(s)}$ depends on $s$ only in its last column.

\section{The matrix $\mathcal{N}^{(s,p)}$\label{app:Nsp}}
For $j\ne p$ we define the entries $\mathcal{N}^{(s,p)}_{j,k}$ of the $(a+b)\times(a+b)$ matrix $\mathcal{N}^{(s,p)}$ as

\bea
\mathcal{N}^{(s)}_{j,k} &=& c\,g^{-1}(w_k,\blac)\,g^{-1}(\bmuc,w_k)
\frac{\partial \tau(w_k|\blac,\bmuc)}{\partial\lac_j},\qquad\quad j=1,\dots,a,\quad j\ne p,
\\
\mathcal{N}^{(s)}_{a+j,k} &=& -c\,g^{-1}(\bmub,w_k )\,g^{-1}(w_k,\blab)
\frac{\partial \tau(w_k|\blab,\bmub)}{\partial\mub_j},\qquad j=1,\dots,b,\quad j\ne p.
\eea
In these formulas one should set $w_k=\lab_k$ for $k=1,\dots,a$ and $w_{k+a}=\muc_k$ for $k=1,\dots,b$.

The $p$-th row has the following elements
\beq
\mathcal{N}^{(s)}_{p,k}=h(\bmuc,w_k)h(w_k,\blab)Y^{(s)}_k,
\eeq
where again $w_k=\lab_k$ for $k=1,\dots,a$ and $w_{k+a}=\muc_k$ for $k=1,\dots,b$, and
\bea
Y^{(s)}_k &=& c\,(\delta_{s1}-\delta_{s2})+(\delta_{s1}-\delta_{s3})\lab_k\left(1-\frac{f(\bmub,\lab_k)}{f(\bmuc,\lab_k)}\right),\qquad\qquad k=1,\dots,a,\nonu
Y^{(s)}_{a+k} &=& c\,(\delta_{s3}-\delta_{s2})+(\delta_{s1}-\delta_{s3})(\muc_k+c)\left(1-\frac{f(\muc_k,\blac)}{f(\muc_k,\blab)}\right),\qquad k=1,\dots,b.
\eea


\begin{thebibliography}{99} 
%
\bibitem{FadST79} L. D. Faddeev, E. K. Sklyanin and L. A. Takhtajan, \textsl{Quantum Inverse Problem. I},
 Theor. Math. Phys. {\bf 40} (1979) 688--706;\\
 L. D. Faddeev and L. A. Takhtajan, \textsl{The quantum method of the inverse problem and the Heisenberg $XYZ$ model},
Usp. Math. Nauk {\bf 34} (1979) 13;  Russian Math. Surveys {\bf 34} (1979) 11 (Engl. transl.).
%
\bibitem{BogIK93L}V. E. Korepin, N. M. Bogoliubov,
A. G. Izergin, \textsl{Quantum Inverse Scattering Method and Correlation Functions}, Cambridge: Cambridge Univ.
Press, 1993.
%
\bibitem{FadLH96} L. D. Faddeev, in: Les Houches Lectures \textsl{Quantum Symmetries}, eds A. Connes
et al, North Holland, (1998) 149.
%
 \bibitem{KulR83}
P. P. Kulish, N. Yu. Reshetikhin,
\textsl{Diagonalization of $GL(N)$ invariant transfer matrices and quantum $N$-wave system (Lee model)}, J.~Phys.  {\bf A 16} (1983) L591--L596.
%
\bibitem{TV}
V. Tarasov and A. Varchenko, \textsl{Combinatorial formulae for nested Bethe vector}, 
SIGMA \textbf{9} (2013) 048 and \texttt{arXiv:math.QA/0702277}.

\bibitem{BR1} 
S.~Belliard and E.~Ragoucy, \textsl{The nested Bethe ansatz for 'all' closed spin chains}, 
J. Phys. \textbf{A41} (2008) 295202, \texttt{arXiv:0804.2822v2}.

\bibitem{BelPRS12b} S. Belliard, S. Pakuliak, E. Ragoucy, N. A. Slavnov,
\textsl{The algebraic Bethe ansatz for scalar products in $SU(3)$-invariant integrable
 models}, J. Stat. Mech. \textbf{1210} (2012) P10017, \texttt{arXiv:1207.0956}.
 %
\bibitem{Ize87} A. G. Izergin, \textsl{Partition function of the six-vertex model in a finite volume},
Dokl. Akad. Nauk SSSR {\bf 297} (1987) 331--333;
Sov. Phys. Dokl. {\bf 32} (1987) 878--879 (Engl. transl.).
%
\bibitem{MaiT00} J. M. Maillet, V. Terras, \textsl{On the quantum inverse scattering problem}, Nucl. Phys. {\bf B 575} (2000) 627--644, \texttt{hep-th/9911030}.
%
\bibitem{BelPRS12c} S. Belliard, S. Pakuliak, E. Ragoucy, N. A. Slavnov,
\textsl{Bethe vectors of $GL(3)$-invariant integrable models}, J. Stat. Mech. \textbf{1302} (2013) P02020, \texttt{arXiv:1210.0768}.
%
\bibitem{Res86}  N. Yu. Reshetikhin, \textsl{Calculation of the norm of Bethe vectors in models with $SU(3)$-symmetry}, Zap. Nauchn. Sem. LOMI {\bf 150} (1986) 196--213;    J. Math. Sci. {\bf 46} (1989) 1694--1706 (Engl. transl.).
%
\bibitem{IzKo} V. E. Korepin, \textsl{Calculation of norms of Bethe wave functions}, Comm. Math. Phys. {\bf 86} (1982) 391--418;\\
 A. G. Izergin,  V. E. Korepin,
\textsl{The quantum inverse scattering method approach to correlation functions},
Comm. Math. Phys. {\bf 94} (1984), 67--92.
%
\bibitem{Whe12} M. Wheeler, \textsl{Scalar products in generalized models with $SU(3)$-symmetry},
Comm. Math. Phys. to appear, \texttt{arXiv:1204.2089}.
%
\bibitem{BelPRS12a} S. Belliard, S. Pakuliak, E. Ragoucy, N. A. Slavnov,
\textsl{Highest coefficient of scalar products in $SU(3)$-invariant models}, J. Stat. Mech. \textbf{1209} (2012) P09003, \texttt{arXiv:1206.4931}.
%
\bibitem{Whe13} M. Wheeler, 
\textsl{Multiple integral formulae for the scalar product of on-shell and off-shell Bethe vectors in SU(3)-invariant models}, Nucl. Phys. {\bf B 875} (2013) 186--212,
\texttt{arXiv:1306.0552}.
%
\bibitem{BelPRS13a} S. Belliard, S. Pakuliak, E. Ragoucy, N. A. Slavnov,
\textsl{Form factors in  $SU(3)$-invariant integrable models}, J. Stat. Mech. \textbf{1309} (2013) P04033, \texttt{arXiv:1211.3968}.
%
\bibitem{PRS13b} S. Pakuliak, E. Ragoucy, N. A. Slavnov,
\textsl{Form factors in quantum integrable models with $GL(3)$-invariant $R$-matrix}, \texttt{arXiv:1312.1488}.
%
\bibitem{BelPRS13d} S. Belliard, S. Pakuliak, E. Ragoucy, N. A. Slavnov, 
\textsl{Bethe vectors of quantum integrable models with $GL(3)$ trigonometric $R$-matrix}, 
SIGMA \textbf{9} (2013) 058, \texttt{arXiv:1304.7602}.

\bibitem{PRS13c} S. Pakuliak, E. Ragoucy, N. A. Slavnov,
\textsl{Scalar products in  models with $GL(3)$ trigonometric $R$-matrix. Highest coefficient},
\texttt{arXiv:1311.3500}.

\bibitem{PRS14a} S. Pakuliak, E. Ragoucy, N. A. Slavnov, 
 in preparation.

\bibitem{KhEOPS} 
 B. Enriquez, S. Khoroshkin, S. Pakuliak, {\sl Weight functions and Drinfeld currents,} \texttt{arXiv:math/0610398},
{Comm. Math. Phys.} {\bf 276} (2007) 691--725;\\
S. Khoroshkin, S. Pakuliak, {\sl  Weight function for $\cU_q(\widehat{\mathfrak{sl}}_3)$,} \texttt{arXiv:math/0610433},
Theor. Math. Phys. {\bf 145} (2005) 1373--1399;\\
S. Khoroshkin, S. Pakuliak, {\sl A computation of an universal weight function for
the quantum affine algebra $\cU_q(\mathfrak{gl}_N)$,} \texttt{arXiv:0711.2819},
J. Math. of Kyoto University {\bf 48} (2008) 277--321;\\
 A. Os'kin, S. Pakuliak, A. Silantyev, {\sl On the universal weight function
for the quantum affine algebra $\cU_q(\mathfrak{gl}_N)$,} \texttt{arXiv:0711.2821},
Algebra and Analysis  {\bf 21} (2009)   196--240;\\
L. Frappat, S. Khoroshkin, S. Pakuliak, E. Ragoucy,
\textsl{Bethe Ansatz for the Universal Weight Function,} \texttt{arXiv:0810.3135},
Ann. H. Poincar\'e \textbf{10} (2009) 513;\\
S. Belliard, S. Pakuliak, E. Ragoucy, 
\textsl{Bethe Ansatz and Bethe Vectors Scalar Products,} \texttt{arXiv:1012.1455},
SIGMA \textbf{6} (2010) 094.

\bibitem{PRS13e} S. Pakuliak, E. Ragoucy, N. A. Slavnov, 
\textsl{Bethe vectors of quantum integrable models based on $\cU_q(\mathfrak{gl}_N)$}, J. Phys. \textbf{A} to appear,
\texttt{arXiv:1310.3253}.

\end{thebibliography}
\end{document}